\documentclass[a4paper]{jpconf}

\usepackage{amssymb,bm}
\usepackage{amsmath,graphicx,color,epsfig}
\usepackage{slashed}
\usepackage{hhline}

\begin{document}

\title{Ward identities, $\bm{ B\to V}$ transition form factors and applications}

\author{M.~Ali Paracha}
\address{Laborat\'orio de F\'isica Te\'orica e Computacional, Universidade Cruzeiro do Sul, 01506-000  S\~ao Paulo SP, Brazil; 
 Department of Physics, School of Natural Sciences, National University of Science and Technology, Islamabad, Pakistan}
\ead{paracha@phys.qau.edu.pk}

\author{Bruno El-Bennich}
\address{Laborat\'orio de F\'isica Te\'orica e Computacional, Universidade Cruzeiro do Sul, 01506-000 S\~ao Paulo, SP, Brazil;
Instituto de F\'isica Te\'orica, Universidade Estadual Paulista, 01140-070 S\~ao Paulo, SP, Brazil }
\ead{bruno.bennich@cruzeirodosul.edu.br}

\author{M.~Jamil Aslam}
\address{Department of Physics, Quaid-i-Azam University, Islamabad 45320, Pakistan}
\ead{jamil@phys.qau.edu.pk}

\author{Ishtiaq Ahmed}
\address{Laborat\'orio de F\'isica Te\'orica e Computacional, Universidade Cruzeiro do Sul, 01506-000 S\~ao Paulo, SP, Brazil;
 National Centre for Physics, Quaid-i-Azam University Campus, Islamabad 45320, Pakistan }
\ead{ishitaq@ncp.edu.pk}

\date{\today}

\begin{abstract}
Long distance effects are studied in the rare exclusive semileptonic $B_{(d,s)}\to V \ell^+ \ell^-$ decays, where $V$ 
denotes a $K^*$ or  $\phi$ meson. The form factors, which describe the meson transition amplitudes in the effective 
Hamiltonian approach, are calculated by means of Ward identities, experimental constraints and
extrapolated within a general vector meson dominance framework. These form factors are then compared to 
the ones obtained in Lattice QCD simulations, with Light Cone Sum Rules and a Dyson-Schwinger equation approach. 
Additionally, the $B_d\to K^* \ell^+ \ell^-$ and $B_s\to \phi  \ell^+ \ell^- $ branching ratios are computed and the differential 
branching fractions are given as a function of the squared-momentum transfer. 
\end{abstract}

\section{Introduction}

Rare $B$ meson decays have been a subject of great interest for the past two decades. These decays not only provide a
stringent tests of the Standard Model (SM), but also serve as a tool to extract physics beyond the SM in the flavor sector. 
We remind that in the SM rare decays do not occur at tree order but proceed at loop level through the Glashow-Iliopoulos-Maiani 
(GIM) mechanism~\cite{GIM}. Attempts to unravel the imprints of physics in and beyond the SM in the flavor sector imply the study 
of inclusive~\cite{incl} and exclusive~\cite{excl} $B$-meson decays. From an experimental point of view, exclusive decays are
easier to observe than inclusive decays. However, theoretically, exclusive decays are more challenging due to the uncertainties
in the calculation of hadronic transition form factors which, so far, are model-dependent quantities when computed 
for the entire range of squared momentum transfer $q^2$. Different frameworks have been used to compute transition 
form factors, namely within the constituent quark model (CQM), light cone sum rules (LCSR), QCD sum rules (QCDSR),
Dyson-Schwinger equation (DSE) approaches and lattice QCD (LQCD) amongst others. These form factors are the ingredients of 
physical observables, such as branching ratios and different asymmetries related to final particle states, especially in 
$B\to K(K^{\ast})\ell^{+}\ell^{-}$~\cite{AAli,Aliev,AAli1}, $B\to\gamma\ell^{+}\ell^{-}$~\cite{SRC} decays.

Nowadays, rare decays, and in particular the decays $B_{d}\to K^{\ast}\mu^{+}\mu^{-}$ and $B_{s}\to\phi\mu^{+}\mu^{-}$, are 
of particular interest as they are the object of intense experimental scrutiny  by the LHCb Collaboration~\cite{LHCb}. Updated values 
of experimental branching ratios for these decays  
have been published~\cite{pdg} and their numerical values are given by,
\begin{eqnarray}
 \mathrm{Br} (B_{d}\to K^{\ast}\mu^{+}\mu^{-})&=&(1.06\pm 0.09)\times 10^{-6}\label{a}, \\
 \mathrm{Br} (B_{s}\to \phi\mu^{+}\mu^{-})&=& (7.6\pm 1.5)\times 10^{-7} \ .\label{b}
\end{eqnarray}
Here, we study the transition form factors, $B_{d}\to K^{\ast}$ and $B_{s}\to \phi$, using Ward identities to compute their values at $q^2=0$ 
and then extrapolate them in a general vector meson dominance model to larger $q^2$ values. We compare these form factors with  
those obtained in Refs.~\cite{AAli,lat,lcsr}. In addition, for the decay $B_{d}\to K^{\ast}\mu^{+}\mu^{-}$, we also compare  
the transition form factors with the results of a DSE model of QCD~\cite{DSE}.
 
The vector dominance approach has been successfully applied to different  heavy-to-light semileptonic decays, such as 
$B\to\rho$~\cite{RD1}, $B\to\gamma$ \cite{RD2}, $B\to K_{1}$\cite{AP} and $B_{c}\to D_{s}^{\ast}$ \cite{AP1}. 
The main purpose here  is to apply the same technique to $B \to K^*(\phi)$ decays, where we relate the form various factors 
of the transition in a model-independent way via Ward identities. This allows us to make a clear separation between 
pole and non-pole type contributions~\cite{JC,JAS}.  The residue of the pole is then determined in a self-consistent 
manner  in terms of $\xi_{\perp}(0)$ or $g_{+}(0)$ which in turn defines  the couplings of vector and axialvector 
$B$ mesons in the $B \to V$ channel. The form factors in the $q^2\to 0$ limit are expressed in terms 
of  a universal function, $\xi_{\perp}(0)\equiv g_{+}(0)$, that is introduced in Large Energy Effective Theory (LEET) ~\cite{JC,JAS} 
of heavy-to-light form factors, and the masses of the particles. The value of $g_{+}(0)$ will be extracted from the experimental
values of the branching ratios of corresponding $B_{d,s} \to V \gamma$ decays. Eventually, using the above form factors, 
we compute branching ratios and compare them with their experimental values in Eqs.~(\ref{a}) and (\ref{b}) as well 
as with the predictions of the LCSR, Lattice QCD and DSE approaches.

 \section{Theoretical Framework}
 \subsection{Effective Hamiltonian}

 At the quark level, the decays $B\to V\ell^{+}\ell^{-}$ ($V=K^{\ast},\phi$) are governed by the transition $b\to s\ell^{+}\ell^{-}$.
 The phenomenology of the rare decays is commonly described  by the heavy-quark effective Hamiltonian,
 \begin{eqnarray}
  H_\mathrm{eff}=-\frac{4G_{F}}{\sqrt{2}}V_{tb}V^{\ast}_{ts} \, \sum\limits_{i=1}^{10} C_{i}({\mu })O_{i}({\mu })\ ,
\label{effham1}
 \end{eqnarray}
where the Wilson coefficients $C_{i}(\mu)$ encode contributions from energy scales above the renormalization point $\mu$; 
due to asymptotic freedom property in QCD,  these can be  computed in perturbation theory, for instance in the naive dimensional
regularization scheme~\cite{Buras}. The local quark operators,  $O_{i}$,  describe via their matrix elements the strong and 
electromagnetic contributions  from {\em all\/} scales below $\mu$ and thus require a nonperturbative approach to their determination. Additional SM/NP  operators can be inserted in the above effective Hamiltonian. The operators which describe 
the decay $B\to V\ell^{+}\ell^{-}$ in the SM are given by,
\begin{eqnarray}
O_{7} &=&\frac{e^{2}}{16\pi ^{2}}m_{b}\left( \bar{s}\sigma _{\mu \nu }P_{R}b\right) F^{\mu \nu },\,  \notag \\
O_{9} &=&\frac{e^{2}}{16\pi ^{2}}(\bar{s}\gamma _{\mu }P_{L}b)(\bar{l}\gamma^{\mu }l),\,  \label{op-form} \\
O_{10} &=&\frac{e^{2}}{16\pi ^{2}}(\bar{s}\gamma _{\mu }P_{L}b)(\bar{l} \gamma ^{\mu }\gamma _{5} l),  \notag
\end{eqnarray}%
where $P_{L,R}=\left( 1\mp \gamma _{5}\right) /2$ are left- and right-handed projectors. The effective Hamiltonian given in
Eq. (\ref{effham1}) yields the following matrix element for the decay $B\to V\ell^{+}\ell^{-}$,
\begin{eqnarray}
 \mathcal{M}(B\rightarrow V \ell^{+}\ell^{-}) &=& \frac{\alpha_\mathrm{em} G_F}{2\sqrt{2}\pi} V_{tb}V_{ts}^{\ast}
 \bigg[  \langle V(k,\varepsilon)| \overline{s}\gamma^{\mu}(1-\gamma^5)b | B(p) \rangle
 \left\{C_9^\mathrm{eff}(\overline{l}\gamma^{\mu}l)+C_{10}(\overline{l}\gamma^{\mu}\gamma^5l)\right\}\notag \\
 &  &  -2\, C_7^\mathrm{eff}m_b\langle
       V(k,\varepsilon)|\overline{s}\, i\sigma_{\mu\nu}\tfrac{q^{\nu}}{q^2}(1+\gamma^5)b|B(p)\rangle(\overline{l}\gamma^{\mu}l)\bigg],
\label{Amplitude}
\end{eqnarray}
where $\alpha_\mathrm{em}$ is the electromagnetic coupling constant while $p$ and $k$ are the momenta of $B$ and $V$ mesons, 
respectively. Moreover, $\varepsilon$ is the polarization vector of final state vector meson and $q^2=(p-k)^2$ is the squared 
momentum transfer.  The explicit expressions of Wilson Coefficients $C_{7,9}^\mathrm{eff}$ and $C_{10}$ are given in 
Ref.~\cite{Wcoefficients} and we do not reproduce them here.

 \subsection{Form Factors and Ward Identities \label{sec2}}
 
Both decays, $B_d\to K^{\ast}\ell^{+}\ell^{-}$ and $B_{s}\to \phi\ell^{+}\ell^{-}$, involve the hadronic matrix elements of the
quark operators introduced in Eq.~(\ref{op-form}) between the $B$ and $V$ ($V=K^{\ast},\phi$) meson states. These matrix 
elements can be parameterized in terms of transition form factors which are functions of the square of momentum transfer, 
$q^{2}$. The different matrix elements can be written as
\begin{eqnarray}
 \left\langle V(k,\varepsilon )\left\vert \bar{s}\gamma_{\mu }b\right\vert B(p)\right\rangle &=&\frac{2\epsilon _{\mu
 \nu \alpha \beta }}{M_{B}+M_{V}}\varepsilon ^{\ast \nu }p^{\alpha }k^{\beta }V(q^{2}),  \label{10a} \\
 \left\langle V(k,\varepsilon )\left\vert \bar{s}\gamma_{\mu }\gamma _{5}b\right\vert B(p)\right\rangle 
&=&  i  \left( M_{B}+M_{V}\right) \varepsilon ^{\ast \mu} A_{1}(q^{2})
\notag \\
& &  -i\, \frac{(\varepsilon ^{\ast }\!\cdot q)}{M_{B}+M_{V}} \left( p+k\right) ^{\mu }A_{2}(q^{2})  \notag \\
& &  -2 i M_V\, \tfrac{q^{\mu }}{q^{2}} \, (\varepsilon ^{\ast }\!\cdot q) \left[ A_{3}(q^{2})-A_{0}(q^{2})\right],   
\label{11}
\end{eqnarray}
where
\begin{eqnarray}
 A_{3}(q^{2})&=&\frac{M_B+M_V}{2M_V}A_{1}(q^{2})-\frac{M_B-M_V}{2M_V}A_{2}(q^{2})\label{A3},
\end{eqnarray}
and where the relation  $A_3(0)=A_0(0)$ holds. Likewise,
\begin{eqnarray}
 \left \langle  V(k,\varepsilon )\left\vert \bar{s}\sigma_{\mu \nu }q^{\nu }b\right\vert B(p)\right\rangle
  & = &   2\,F_{1}(q^{2})\epsilon _{\mu\nu \alpha \beta }\, \varepsilon ^{\ast \nu }p^{\alpha }k^{\beta } \label{F1}  \\
 \left \langle  V(k,\varepsilon )\left\vert \bar{s}\sigma_{\mu \nu }q^{\nu }\gamma ^{5}b\right\vert B(p)\right\rangle
  & = &  i \left[ \!\!
    \begin{array}{c} 
      \left( M_{B}^{2}-M_{V}^{2}\right) \varepsilon _{\mu}^{\ast }-(\varepsilon ^{\ast }\cdot q)(p+k)_{\mu }
     \end{array}%
            \!\!   \right] F_{2}(q^{2})      \notag \\
  &  + &   i \,  (\varepsilon ^{\ast }\cdot q)\left[ \!
   \begin{array}{c}
       q_{\mu } - \frac{q^{2}}{M_{B}^{2}-M_{V}^{2}}(p+k)_{\mu }%
   \end{array}%
           \! \right] F_{3}(q^{2}) \ ,
  \label{F3}
\end{eqnarray}
where $F_{1}(0)=F_{2}(0)$. Now, the form factors appearing in Eqs.~(\ref{10a}) to (\ref{F3}) can be related to each other by means
of  Ward identities~\cite{RD1,RD3}, i.e.,
\begin{eqnarray}
 \left\langle V(k,\varepsilon )\left\vert \bar{s}\sigma_{\mu \nu }q^{\nu }b\right\vert B(p)\right\rangle = 
  -(m_{b}+m_{s})\left\langle V(k,\varepsilon )\left\vert \bar{s}\gamma_{\mu }b\right\vert B(p)\right\rangle\label{WI1}   \\
    \left\langle V(k,\varepsilon )\left\vert \bar{s}\sigma_{\mu \nu }q^{\nu }\gamma ^{5}b\right\vert B(p)\right\rangle 
    = (m_{b}-m_{s})\left\langle  V(k,\varepsilon )\left\vert \bar{s}\gamma_{\mu }\gamma _{5}b\right\vert  B(p)\right\rangle\label{WI2} \ .
\end{eqnarray}
Making use of  Eqs.~(\ref{10a}) to (\ref{F3}) in Eqs.~(\ref{WI1}) and (\ref{WI2}), one can relate the transition form factors as follows:
\begin{eqnarray}
 F_{1}(q^{2})&=&\frac{m_{b}+m_{s}}{M_{B}+M_{V}}\, V(q^{2})\label{WI3},\\
 F_{2}(q^{2})&=&\frac{m_{b}-m_{s}}{M_{B}-M_{V}}\,  A_{1}(q^{2})\label{WI4},\\
 F_{3}(q^{2})&=&-(m_{b}-m_{s})\tfrac{2M_{V}}{q^{2}}[A_{3}(q^{2})-A_{0}(q^{2})] \ .
 \label{WI5}
\end{eqnarray}

Note that the form factor relations as stated in Eqs.~(\ref{WI3}), (\ref{WI4}) and (\ref{WI5}) are model independent. Yet, in deriving the 
above Ward identities~\cite{RD1}, one assumes that the light-quark degrees of freedom can be neglected and hence takes 
$p=p_b$ as suggested by heavy quark effective theory in the limit $m_b \to \infty$ with $p_b/m_b$ constant. This is a 
sensible approach and justified  in the case of $B$ mesons, though nonperturbative $\Lambda_\mathrm{QCD}/m_c$ corrections 
due to the light quarks may be important in charmed mesons. As known from a series of nonperturbative studies, heavy quark effective 
theory is not generally a good guide to charm physics~\cite{ElBennich:2009vx,ElBennich:2012tp,ElBennich:2011py,
Bashir:2012fs,Rojas:2014aka} since the charm quark is neither light nor really heavy. Eventually, a reliable calculation of 
heavy-light form factors, couplings and decay constants requires the correct description of dynamical chiral symmetry breaking, 
that is the effect of the light degrees of freedom.

\begin{figure}[t]
\centering
  \includegraphics[scale=.8]{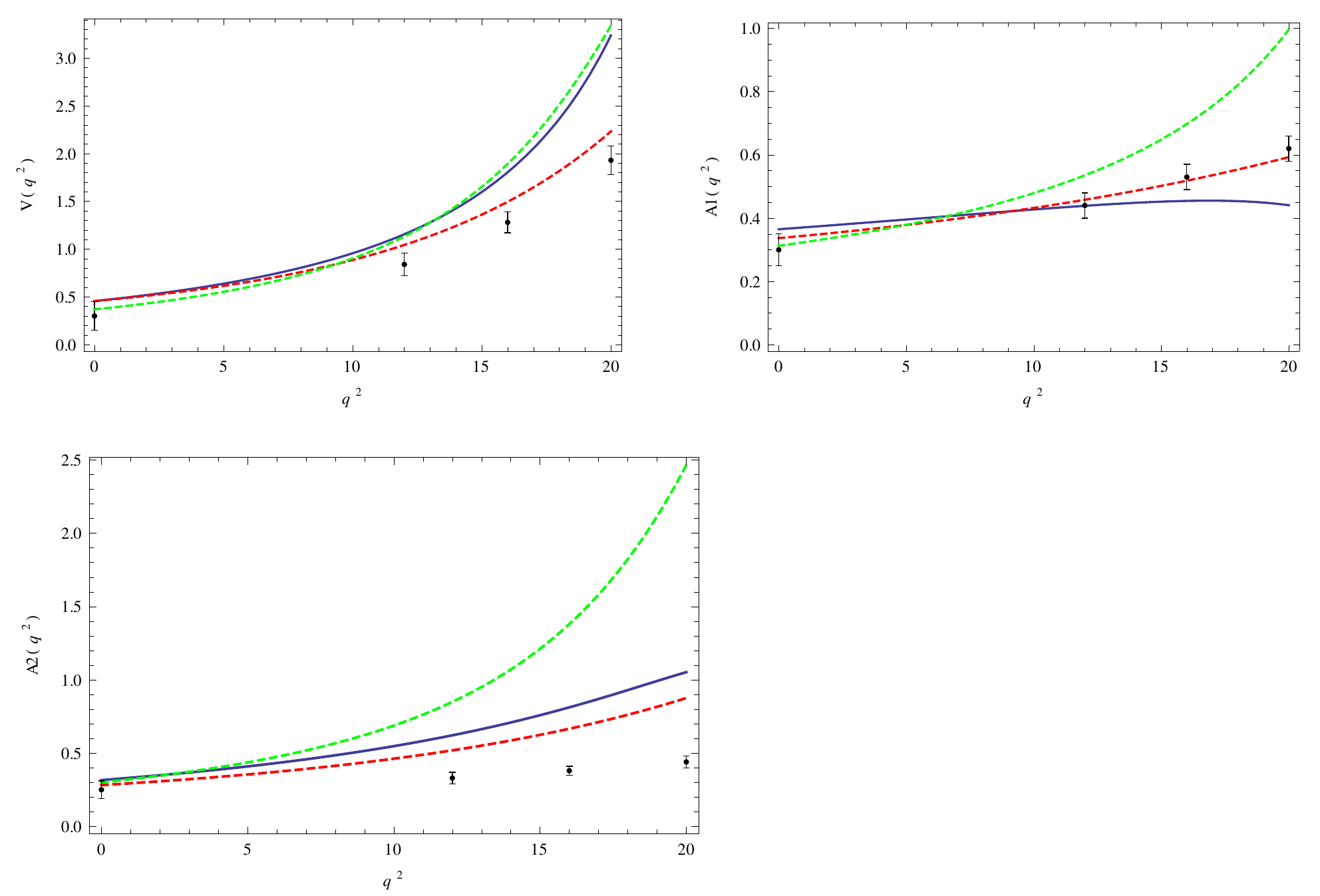} 
\caption{Comparison of the $B_{d}\to K^\ast$ transition form factors, Eqs.~(\ref{V}), (\ref{A1}) and (\ref{A2}) (black solid curve) with predictions 
from LQCD~\cite{lat} (points with error bars),  LCSR~\cite{AAli} (red dashed curve) and DSE~\cite{DSE} (green dashed curve).}
  \label{fig1}
\end{figure}
\begin{figure}[t]
\centering
  \includegraphics[scale=.8]{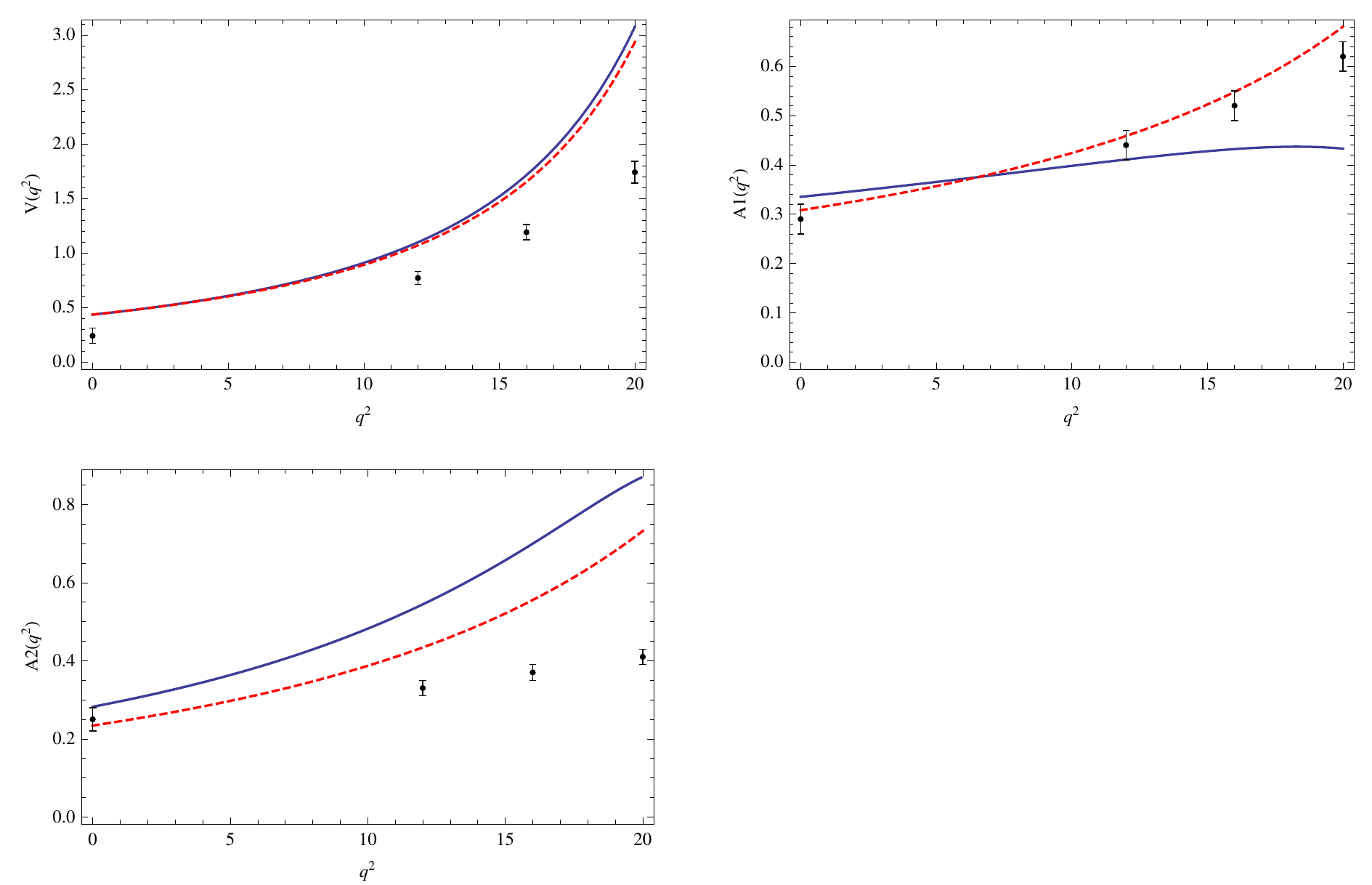} 
\caption{Comparison of form factors for $B_{s}\to \phi$; curve description as in Fig.~\ref{fig1}
though no DSE results available in this transition.} 
  \label{fig2}
\end{figure}

The normalization of the form factors  at $q^{2}=0$ can be expressed in  terms of $g_{+}(q^2), g_{-}(q^2)$, $h(q^2)$ and $h_{1}(q^2)$ 
which are discussed in detail in Refs.~\cite{RD1,AP1}. Namely,  we me make use of
\begin{eqnarray}
\langle V(k,\varepsilon^{\ast})|\bar s i\sigma_{\alpha\beta}b|B(p)\rangle &=& 
     - i\, \epsilon_{\alpha\beta\rho\sigma}\varepsilon^{\ast\rho}  
    [(p+k)^{\sigma}g_{+} + q^{\sigma}g_{-}]-(\varepsilon^{\ast}\!\cdot q)\, \epsilon_{\alpha\beta\rho\sigma}(p+k)^{\rho}q^{\sigma}h
    \notag \label{def}\\
& &  -i  [ (p+k)^{\alpha}\epsilon_{\beta\rho\sigma\tau}\varepsilon^{\ast\rho}(p+k)^{\sigma}q^{\tau}-\alpha\leftrightarrow\beta]\, h_{1} \ ,
\end{eqnarray}
so the form factors $V(q^2), A_{1,2}(q^2), F_{1,2,3}(q^2)$ become,
\begin{eqnarray}
 F_{1}(q^{2})&=&  g_{+}(q^{2})-q^{2}h(q^{2}) \label{N1} \ ,\\
 F_{2}(q^{2})&=&g_{+}(q^{2})+\frac{q^{2}}{M^{2}_{B}-M^{2}_{V}}\, g_{-}(q^{2})\ ,   \label{N2}\\
 F_{3}(q^{2})&=&-g_{-}(q^{2})-(M^{2}_{B}-M^{2}_{V})h(q^{2}) \ ,   \label{N3}
 \end{eqnarray}
 and
 \begin{eqnarray}
 V(q^{2})&=&\frac{M_{B}+M_{V}}{m_{b}+m_{s}}\left [ \, g_{+}(q^{2})-q^{2}h_{1}(q^{2})  \right ] \ ,   \label{N4}\\
 A_{1}(q^{2})&=&\frac{M_{B}-M_{V}}{m_{b}-m_{s}}\left [g_{+}(q^{2})+\frac{q^{2}}{M^{2}_{B}-M^{2}_{V}}\, g_{-}(q^{2})\right]\  ,  \label{N5}\\
 A_{2}(q^{2})&=&\frac{M_{B}-M_{V}}{m_{b}-m_{s}}\, \left [ g_{+}(q^{2})-q^{2}h(q^{2})\right ]-\frac{2M_{V}}{M_{B}-M_{V}}\, A_{0}(q^{2})\ . \label{N6}
\end{eqnarray}
 At $q^{2}=0$, the form factors $F_{1}$, $F_{2}$, $V$ and $A_{1}$ in Eqs.~(\ref{N1}), (\ref{N2}), (\ref{N4}) and (\ref{N5}) are thus 
 parameterized by a single constant $g_{+}(0)$ whereas $A_{2}$ and $A_{0}$  in Eq.~(\ref{N6}) can be expressed in terms 
 of $g_{+}(0)$ and $A_{0}(0)$. It is well known that in case of real photon emission 
 in $B_{d,s}\to V\gamma$ decays, the decay rate 
 is a function of  $g_{+}(0)$~\cite{JAS}:
\begin{eqnarray}
 \Gamma(B\to V\gamma)=\frac{G_{F}^{2}\, \alpha_\mathrm{em}}{32\pi^{4}}|V_{tb}V_{ts}^{\ast}|^{2}m^{2}_{b}M^{3}_{B}
 \left ( 1-\tfrac{M^{2}_{V}}{M^{2}_{B}} \right )^{\!3}|C_{7}^\mathrm{eff}|^{2}|g_{+}(0)|^{2}.
\end{eqnarray}
Using the experimental values of branching ratios of $B_{d}\to K^{\ast}\gamma$ and $B_{s}\to\phi\gamma$ decays~\cite{pdg},
\begin{eqnarray}
  \mathrm{Br} (B_{d}\to K^{\ast}\gamma)&=&(4.33\pm 0.15)\times 10^{-5}\label{ra1}, \\
 \mathrm{Br} (B_{s}\to \phi\gamma)&=& (3.6\pm 0.4)\times 10^{-5} \ ,\label{rb1}
\end{eqnarray}
the extracted $g_{+}(0)$ for these decays are, 
\begin{eqnarray}
 g_{+}(0)^{B_{d}\to K^{\ast}}&=&0.365^{+0.025}_{-0.025},\label{gK}\\
 g_{+}(0)^{B_{s}\to\phi}&=&0.335^{+0.02}_{-0.02} \  .
 \label{gphi}
\end{eqnarray}
With $g_{+}(0)$ in hand, the other unknown, i.e. $A_0(0)$, can be expressed in terms of it as~\cite{RD1},
 \begin{eqnarray}
 A_0(0) & = & \left ( \frac{1-M_{V}^2/M_{B}^2}{1+M_{V}^2/M_{B}^2}+\frac{M_B}{M_V} \right  )g_{+}(0)\ .
 \end{eqnarray}

 It is worth mentioning that the form factor relations derived from Ward identities do not hold for the entire physical  momentum, $q^{2}$, 
 region. Therefore, we employ the following parametrization between $q^{2}=0$ and near the poles:
 \begin{eqnarray}
     F(q^{2})  & = &    \frac{F(0)}{(1-q^{2}/M^{2})(1-q^{2}/M^{\prime 2})}\ ,  
     \label{P1}
 \end{eqnarray}
where $M$ is $M_{B^{\ast}} (1^-)$ or $M_{B^{\ast}_A}(1^+)$ and $M^{\prime}$ is the radial excitation of $M$. The parametrization given 
in Eq.~(\ref{P1}) not only takes into account the correction to the single pole dominance, as suggested by dispersion relations~\cite{RD3}, 
but also help us to determine the couplings of $B^{\ast}$ or $B^{\ast}_{A}$ to the $BV$ channel~\cite{RD1,RD2}, which is not the aim of this work. 
Finally, using the above parameterization, the form factors can be expressed as:
 \begin{eqnarray}
  V(q^{2})&=&\frac{V(0)}{(1-q^{2}/M^{2}_{B^{\ast}})(1-q^{2}/M^{\prime2}_{B^{\ast}})},   \label{V} \\
   A_{1}(q^{2})&=&\frac{A_{1}(0)}{(1-q^{2}/M^{2}_{B^{\ast}})(1-q^{2}/M^{\prime2}_{B^{\ast}})}\left ( 1-\frac{q^{2}}{M^{2}_{B} -M^{2}_{V}} \right ),\label{A1}    \\
   A_{2}(q^{2})&=&\frac{{\tilde A_{2}}(0)}{(1-q^{2}/M^{2}_{B^{\ast}})(1-q^{2}/M^{\prime 2}_{B^{\ast}})}
  -\frac{2M_{V}}{M_{B}-M_{V}}\frac{A_{0}(0)}{(1-q^{2}/M^{2}_{B})(1-q^{2}/M^{\prime2}_{B})},\label{A2}
 \end{eqnarray}
 where $\tilde A_{2}(0)$ is defined as \cite{RD1}:
 \begin{eqnarray}
  \tilde A_{2}(0)& = & \left (\frac{M_{B}-M_{V}}{m_{b}-m_{s}}\right )\, g_{+}(0) \ .
 \end{eqnarray}

The transition form factors  in Eqs.~(\ref{V}), (\ref{A1}) and (\ref{A2}) are plotted as a function of $q^2$ in Figs.~\ref{fig1} and \ref{fig2} for $B_{d}\to K^{\ast}$ 
and  $B_{s}\to \phi$, respectively, where we use the Particle Data Group values for the masses of $B$, $B^*$, $B_A^*$ and their excited states~\cite{pdg}. 
In the same figures, we also compare our form factors with those obtained in the Lattice QCD and LCSR approaches and in case of the $B_{d}\to K^{\ast}$ 
transition the DSE form factors~\cite{DSE} are also included. The trend of the $V(q^2)$ evolution we obtain parallels that of the DSE form factor, 
whereas our $A_1(q^2)$ and $A_2(q^2)$ compare more favorably with the LCSR results, as becomes clear from Fig.~\ref{fig1}. Similarly, for the 
$B_{s} \to \phi \mu^+ \mu^-$ decay, our form factors are mostly in agreement with LCSR predictions except for larger momenta, $q^2 \gtrsim 10$~GeV$^2$; 
see Fig.~\ref{fig2}. It is also apparent that lattice QCD simulations produce a softer slope of the $A_1(q^2)$ and $V(q^2)$ form factors than the other calculations. 
For reference, we also list the form factor values at $q^2=0$ and $q^{2}=q^{2}_\mathrm{max}$  as well as their ratios in Tables~\ref{tab1} and \ref{tab2}.

\begin{table}[t]
\begin{tabular}{|c|c|c|c|c|c|c|}
\hline
\hline
 & $V$&$A_{1}$ & $A_{2}$ & $\frac{V(q^{2}_\mathrm{max})}{V(0)}$ & $\frac{A_{1}(q^{2}_\mathrm{max})}{A_{1}(0)}$ & $\frac{A_{2}(q^{2}_\mathrm{max})}{A_{2}(0)}$ 
 \\ \hline
Present Work & 0.456\ (2.858) & 0.365 \ (0.4476)& 0.316 \ (1.007) & 6.267 & 1.226 & 3.186\\ \hline
LCSR & 0.457 \ (2.064) & 0.337 \ (0.577) & 0.282\ (0.830)& 4.516 & 1.712 & 2.943\\ \hline
LQCD & 0.30 \ (1.93) &  0.30 \ (0.62) & 0.25 \ (0.44)& 6.433 & 2.06 & 1.76\\\hline
DSE & 0.37 \ (2.98) & 0.29 \ (1.28) & 0.30\ (2.177) & 8.054 & 4.431 & 7.256\\\hline
\end{tabular}
 \caption{Variation of form factors for the transition $B_{d}\to K^{\ast}$ at $q^{2}=0$ and $q^{2}=q^{2}_\mathrm{max}$ in different approaches. 
 The first value represents the form factors at $q^2=0$ whereas the value at maximum $q^{2}$ is in parentheses.
 The ratios of the form factors at $q^{2}_\mathrm{max}$ and $q^{2}=0$ are listed in the last three columns.}
\label{tab1} 
\end{table}

\begin{table}[t]
\begin{tabular}{|c|c|c|c|c|c|c|}
\hline
\hline
 & $V$ & $A_{1}$ & $A_{2}$ & $\frac{V(q^{2}_\mathrm{max})}{V(0)}$ & $\frac{A_{1}(q^{2}_\mathrm{max})}{A_{1}(0)}$ & $\frac{A_{2}(q^{2}_{max})}{A_{2}(0)}$ \\\hline
Present Work & 0.433 \ (2.566) & 0.335 \ (0.436) & 0.282 \ (0.828) &5.926 & 1.301 & 2.936\\ \hline
LCSR & 0.434 \ (2.455) & 0.311\ (0.637) & 0.234 \ (0.676) & 5.656 & 2.048 & 2.888 \\ \hline
Lattice & 0.24 \ (1.74) & 0.29 \ (0.62) & 0.25 \ (0.41) & 7.25 & 2.137 & 1.64\\\hline
\end{tabular}
\caption{Variation of form factors for the transition $B_{s}\to\phi$ at $q^{2}=0$ and $q^{2}=q^{2}_\mathrm{max}$ in different approaches. Table entries are as in 
 Table~\ref{tab1}.}
 \label{tab2}
\end{table}

\section{Applications of Transition Form Factors: Branching Fractions}

To conclude, we present the calculated branching ratios of the $B \to K^*\mu^+\mu^-$ and $B_{s}\to \phi \mu^+ \mu^-$ decays, where we remind that 
the transition form factors are the major hadronic input as well as source of uncertainties.  Here we use the form factors introduced in Section~\ref{sec2} and 
extrapolated via Eqs.~(\ref{V}), (\ref{A1}) and (\ref{A2}) to compute the differential branching fractions and compare them to those obtained with 
form factors of the corresponding LCSR, Lattice QCD and DSE approaches. 

\begin{figure}[t]
\includegraphics[scale=0.55]{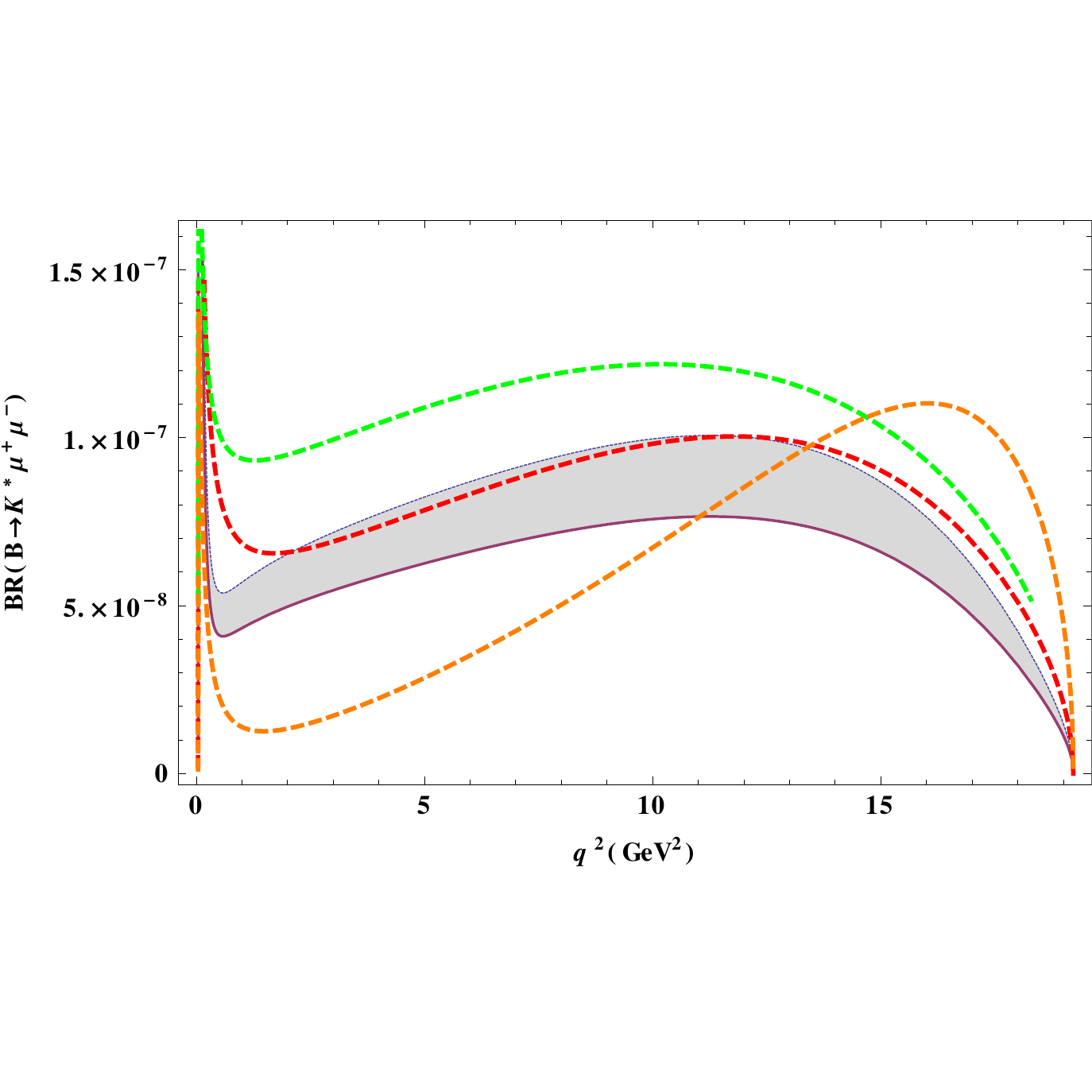}  \hfill
\includegraphics[scale=0.55]{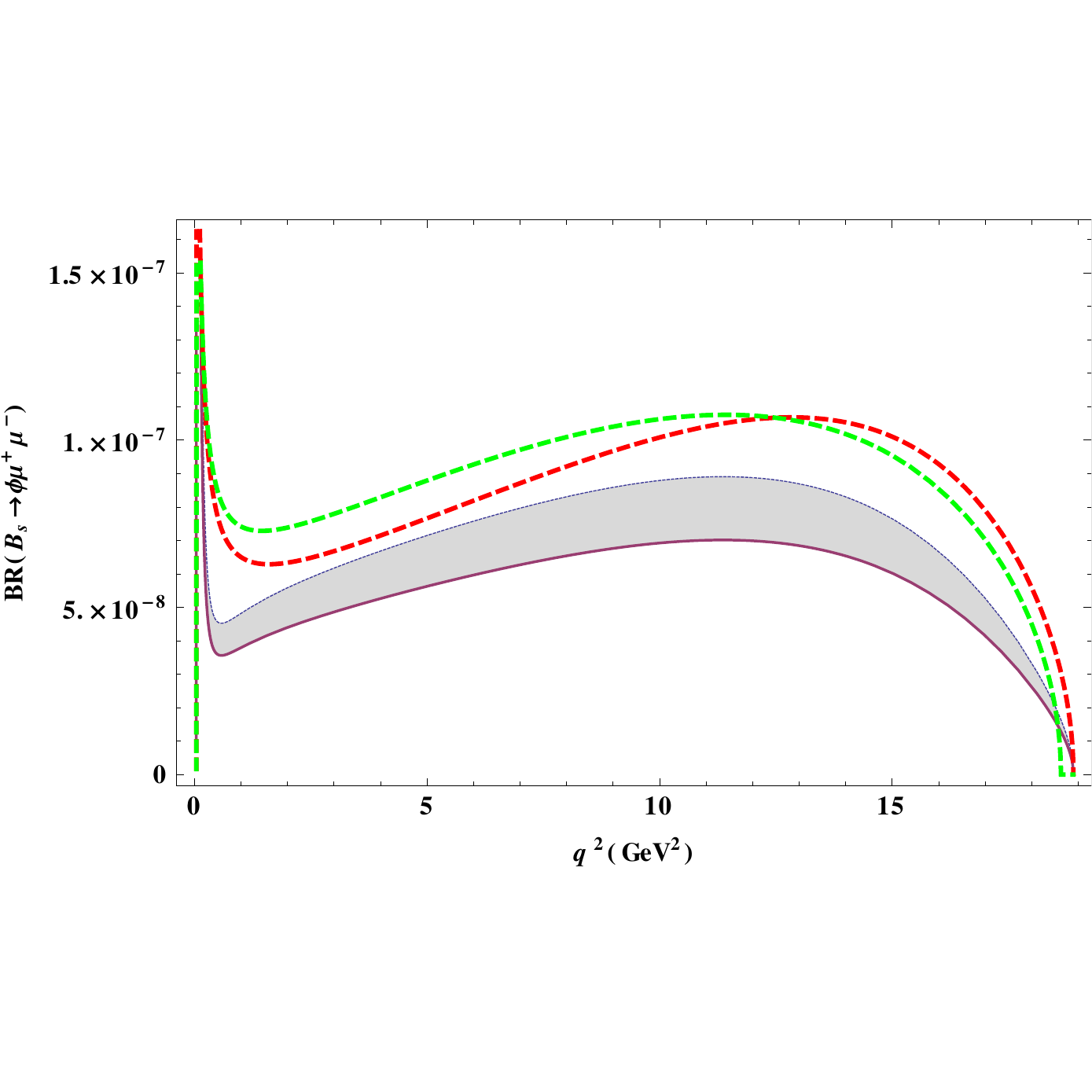} 
\vspace*{-1.2cm}
\caption{Differential branching fractions, $d\Gamma/dq^2$, for the decays $B_{d}\to K^{\ast}\mu^{+}\mu^{-}$ (left panel) and  $B_{s}\to\phi\mu^{+}\mu^{-}$ 
(right panel). The gray-shaded  band represents the estimated error of the differential branching fraction in the present approach; the same branching fractions
computed with the LCSR,  Lattice QCD and DSE form factors are plotted with red-dashed, green-dashed and orange-dashed curves, respectively. }
\label{fig3}
\end{figure}

The formula for the differential decay rate is given by,
\begin{eqnarray}
   \frac{d^{2}\Gamma(B\to V\ell^{+}\ell^{-})}{d\cos\theta\, dq^2}&=&
   \frac{1}{2M^{3}_{V}}\frac{2\beta\sqrt{\lambda}}{(8\pi)^3}| \, \mathcal{M}|^{2}\label{10}
 \end{eqnarray}
with $\mathcal{M}$ from Eq.~(\ref{Amplitude}) and  where $\beta\equiv\sqrt{1- 4m^2_\ell/s}$, $\lambda\equiv \lambda(M_{B},M_{V},s) = M^{4}_{B} +M^{4}_{V}
 +q^{4}-2M^{2}_{{B}}M^{2}_{V}-2q^2M^{2}_{B}-2q^2M^{2}_{V}$. The explicit form of differential decay rates for the decay $B_{d}\to K^{\ast}\mu^{+}\mu^{-}$
 and $B_{s}\to\phi\mu^{+}\mu^{-}$ can be found in Ref.~\cite{AP2}. The plot of the differential branching fractions, $d\Gamma(B\to V\ell^{+}\ell^{-})/dq^2$, 
for the above mentioned decays in different approaches are depicted in Figs.~\ref{fig3}a and \ref{fig3}b. The functional form of the differential branching 
fractions we obtain is comparable with the one using other form factor calculations, albeit with variations in magnitude. Note, however, that the DSE approach 
leads to a stronger variation of $d\Gamma/dq^2$ with $q^2$ in the $B \to K^*$ transition. 

Furthermore, numerical values of branching ratios of $B_{d}\to K^{\ast}\mu^{+}\mu^{-}$ and $B_{s}\to\phi\mu^{+}\mu^{-}$ decays are listed in Table~\ref{tab3}.
The $B_d \to K^\ast \mu^+ \mu^-$ branching ratio obtained with transition form factors of the DSE based model lies close to the experimental value. Within theoretical 
and experimental errors, both of which may be underestimated, our numerical results for both decays do agree with the PDG average. We cannot make a 
statement in case of the  LQCD and LCSR results as no error estimate is given, yet the LQCD branching ratios are considerably larger than experimental values. 
This could  be due to the fact that LQCD does not produce as strongly rising form factors, hence might not describe a VMD behavior. In the case of the 
$B_s \to \phi \mu^+ \mu^-$ decay, the discrepancy between the experimental average and LCSR and LQCD values is equally large. It is worthwhile to remind
that the values of $g_{+}(0)$, Eqs.~(\ref{gK}) and (\ref{gphi}), are very similar. Given that the functional form of the $V(q^2)$, $A_1(q^2)$ and $A_2(q^2)$ 
form factors of the $B_d \to K^\ast$ transition is not noticeable different from the one in $B_s \to \phi$ and that the phase space difference is minor, we do not expect
very different branching ratio values.

 \begin{table}
 \centering
 \begin{tabular}{|c|c|c|c|}
\hline
\hline
 & Br$(B_{d}\to K^{\ast}\mu^{+}\mu^{-})$& Br$(B_{s}\to\phi\mu^{+}\mu^{-})$\\\hline
Present Work & $(1.37\pm 0.2)\times 10^{-6}$ & $(1.21\pm 0.15)\times  10^{-6} $                   \\\hline
  LCSR & $1.31\times 10^{-6}$ & $1.70\times 10^{-6} $                    \\\hline
  Lattice& $1.97\times 10^{-6}$ & $1.67\times 10^{-6}$\\\hline
  DSE& $1.18\times 10^{-6}$ & --- \\\hline
  PDG     & $(1.06\pm 0.09)\times 10^{-6}$ &  $(7.6\pm 1.5)\times 10^{-7}$                      \\\hline
 \end{tabular}
 \label{tab3}
 \caption{Numerical values of branching ratios for the semileptonic decays $B_{d}\to K^{\ast}\mu^{+}\mu^{-}$  and $B_{s}\to\phi\mu^{+}\mu^{-}$ in 
 different form factors approaches.}
 \end{table}

\section*{Acknowledgments}
M. A. P., B. E.  and I. A. are grateful to the organizers of the {\em XXXVII Reuni\~ao de Trabalho sobre F\'isica Nuclear no Brasil\/} in Maresias, S\~ao Paulo,
for their kind invitation and acknowledge support by the S\~ao Paulo Research Foundation (FAPESP) under grant nos.~2012/13047-2, 2013/23177-3 and 
2013/16088-4. M. A. P. and I. A. would also like to thank the Physics Department at Quaid-i-Azam University for the kind hospitality during the last stages of 
the work. M. J. A would like to thank the support by Quaid-i-Azam University through the University Research Fund. B. E. is partially supported by a 
CNPq fellowship no.~301190/2014-3.

\end{document}